\documentclass[pre, twocolumn,showpacs,preprintnumbers,amsmath,amssymb ]{revtex4}% Physical Review Letters

\usepackage{graphicx}% Include figure files
\usepackage{dcolumn}% Align table columns on decimal point
\usepackage{bm}% bold math
\usepackage{float}

\usepackage{color}
\definecolor{qmcgreen}{RGB}{0, 255, 0}

\begin{document}

\input epsf.sty

\newcommand{\Ham}{{\cal H}}
\newcommand{\Ho}{{\cal H}_0}
\newcommand{\Hp}{{\cal H}_{\hbox{\scriptsize pert}}}
\newcommand{\Eav}{E_{\hbox{\scriptsize av}}}
\newcommand{\Nb}{N_{\hbox{\scriptsize b}}}

\title{Quantum quenches with random matrix Hamiltonians and disordered potentials}
\author{Fabian Kolley}
\affiliation{Institute for Theoretical Physics, University of Heidelberg, Philosophenweg 19, 69120 Heidelberg, Germany}
\author{Oriol Bohigas}
\affiliation{CNRS, Universit\'{e} Paris-Sud, UMR8626, LPTMS, Orsay Cedex, F-91405, France}
\author{Boris V. Fine}
\affiliation{Institute for Theoretical Physics, \ University of Heidelberg, Philosophenweg 19, 69120 Heidelberg, Germany}
\email{B.Fine@thphys.uni-heidelberg.de}

\date{September 13, 2012}

\begin{abstract}
We numerically investigate statistical ensembles for the occupations of eigenstates of an isolated quantum system emerging as a result of quantum quenches. The systems investigated are sparse random matrix Hamiltonians and disordered lattices.  In the former case, the quench consists of sudden switching-on the off-diagonal elements of the Hamiltonian. In the latter case, it is sudden switching-on of the hopping between adjacent lattice sites. The quench-induced ensembles are compared with the so-called ``quantum micro-canonical" (QMC) ensemble describing quantum superpositions with fixed energy expectation values. Our main finding is that quantum quenches with sparse random matrices having one special diagonal element lead to the condensation phenomenon predicted for the QMC ensemble. Away from the QMC condensation regime, the overall agreement with the QMC predictions is only qualitative for both random matrices and disordered lattices but with some cases of a very good quantitative agreement.   In the case of disordered lattices, the QMC ensemble can be used to estimate the probability of finding a particle in a localized or delocalized eigenstate.
\end{abstract}

\maketitle

\section{Introduction}\label{sec:introduction}

The issue of statistical ensembles emerging in isolated quantum systems after a sudden perturbation, often referred to as ``quantum quench'', has recently come into a sharp focus of statistical physics research\cite{Kollath-07,Barthel-08,Genway-10,Ponomarev-11,Banuls-11,Ji-11,Ganahl-12,Brandino-12}. Of particular interest for the present paper are the statistical ensembles for the occupations of the energy eigenstates of an isolated quantum system. We consider two kinds of quenches: a more abstract one induced by switching-on the off-diagonal elements of a sparse random matrix Hamiltonian, and a more physical one induced by  switching-on random hopping elements in a disordered potential. Our interest was motivated by the numerical result reported in Ref.~\cite{Fine-09-statistics}, that, for a certain class of random matrices, the above quench induces the so-called ``quantum micro-canonical" (QMC) ensemble \cite{Wootters-90, Brody-98, Bender-05, Fine-09-statistics, Fresch-10A, Fresch-10B}. The QMC ensemble, when treated analytically, exhibits a remarkable phenomenon of condensation obtained in Refs.~\cite{Fine-09-statistics, Fresch-10A, Fresch-10B}. However, the numerical investigation of Ref.~\cite{Fine-09-statistics} could not reach the condensation regime with the class of random matrices considered. 

In this paper, we investigate a new class of random matrices with a special diagonal element, which allows us to reach the QMC condensation regime. At the same time, we show that the emerging ensembles often show noticeable differences from the QMC predictions.  We find that the QMC condensation is connected to the localization properties of the eigenstates near the edge of the energy spectrum.
We also find that the ensembles emerging in quenched disordered potentials are less universal than in the case of quenched random matrices. However, they are still in a qualitative agreement with the QMC predictions. The random matrix quench may possibly be used to mimic the effect of multiple small quenches on physical systems with a large number of quantum levels but not too large a number of particles \cite{Ji-11}.
The disordered-potential quench is relevant to the recent experiments\cite{Schwartz-07,Aspect-08,Roati-08,DeMarco-11,Jendrzejewski-12}, which report the evidence of Anderson localization \cite{Anderson-58}. It can, in particular, help to estimate the probability of a particle being below or above the mobility edge after the quench. 

The rest of the paper is organized as follows: In Sec.~\ref{sec:general}, we give the general formulation of the problem. In Sec. \ref{sec:QMC}, we describe the QMC ensemble. Sections~\ref{sec:RandomMatrices} and \ref{sec:DisorderedPotentials} deal with the random matrix Hamiltonians and the disordered potentials, respectively.

\section{General formulation}
\label{sec:general}

We consider an isolated quantum system having a large but finite number of quantum states $N$ with a Hamiltonian that can be decomposed as follows:
\begin{equation}
{\cal H} = \Ho + \Hp.
\label{eq:H}
\end{equation}
In the context of sparse random matrices, ${\cal H}$ represents the entire random matrix, $\Ho$ the diagonal elements of ${\cal H}$ and $\Hp$  the off-diagonal elements. In the case of disordered lattices, $\Ho$ represents the on-site energies, and $\Hp$ the hopping terms. We denote the eigenstates of ${\cal H}$ as $\psi_i$ and the corresponding eigenenergies as $E_i$. The eigenstates of the Hamiltonian $\Ho$ are to be denoted as $\phi_k$. We also choose the origin of the energy axis such that $\sum_{i=1}^N E_i = 0$, i.e., for a symmetric energy spectrum, $E_i= 0$ corresponds to the middle of the spectrum.

We ask the following question: if the system is initially in an eigenstate $\phi_k$ of the Hamiltonian $\Ho$, and then $\Hp$ is switched on, what kind of statistical ensemble emerges for the occupations of the eigenstates $\psi_i$ of the Hamiltonian ${\cal H}$? Specifically, we consider the expansion 
\begin{equation}
\phi_k = \sum_i c_i \psi_i ,
\label{eq:phik}
\end{equation}
where $c_i$ are the complex amplitudes, and investigate the average values for the occupations of eigenstates $\langle |c_i|^2 \rangle$ as a function of energy $E_i$ and energy $\Eav \equiv \langle \phi_k |  {\cal H} | \phi_k \rangle = \langle \phi_k |  \Ho | \phi_k \rangle $. The latter equality is due to the fact, that, in the both cases considered, $\Hp$ does not have diagonal elements in the eigenbasis of $\Ho$. In other words, the $k$-th diagonal element of $\Ho$ fixes the energy expectation value of the of $\phi_k$ with respect to ${\cal H}$. For convenience, we also introduce the ``occupations of eigenstates'' $p_i \equiv |c_i|^2$. The averaged occupations of eigenstates as a function of their energy is to be denoted as  $\langle p_i \rangle (E_i)$. In the following, we often drop index $i$ in this function and represent it as $\langle p \rangle (E)$.

The quantity $\langle p_i \rangle (E_i)$ is to be obtained numerically by organizing eigenstates $\psi_i$ with close values of $E_i$ into ``energy bins'' and then  by averaging within each bin. The statistics will be further improved by including within a given energy bin all $p_i$s for the ensemble of initial states $\phi_k$ with close values of $\Eav$ and for different random realizations of random matrix Hamiltonians or disordered potentials. 

The present work is complementary to a number of studies that considered the statistics of eigenvectors for band random matrices \cite{Casati-90,Zyczkowski-92,Mirlin-93} and quantum quenches for other special kinds of sparse random matrices \cite{Brandino-12}. The distinctive focus here is on the statistics for a fixed value of $\Eav$.
If one were to perform the above averaging over all initial states $\phi_k$, i.e. not just over those having close values of $\Eav$, this would result in $\langle p \rangle (E) = 1/N$, i.e. there would be no energy dependence of $\langle p \rangle$. However, the constraint of the energy expectation value $\Eav$ is incompatible with the constant value of $\langle p \rangle$. For example, if $\Eav $ is below the middle of the energy spectrum of ${\cal H}$, then it should be expected that $\langle p \rangle (E)$ is a decreasing function of $E$. In general, this decrease is not expected to be exponential. Investigation of the character of this decrease is the goal of the present paper.

\section{The QMC ensemble}\label{sec:QMC}

The QMC ensemble is a quantum ensemble for which the constraint on the energy expectation value is {\it the only} constraint in the Hilbert space of the system \cite{Brody-98, Fine-09-statistics, Fresch-10A}. In the Euclidean space of variables $p_i$, the QMC ensemble is defined by the uniform probability density on the manifold constrained by the conditions
\begin{equation}
 \sum_{i=1}^{N}E_{i}p_{i}=\Eav ,
 \label{Eav}
\end{equation}
\begin{equation}
 \sum_{i=1}^{N}p_{i}=1 ,
 \label{norm}
\end{equation}
and
\begin{equation}
 p_{i}\geq 0 .
 \label{ineq}
\end{equation}
The phases of the complex amplitudes $c_i$ are completely random.

Contrary to naive expectations, the QMC ensemble does not lead to the conventional thermodynamics. The marginal probability distribution $P_{i}(p_{i})$ associated with the QMC ensemble was derived in Ref. \cite{Fine-09-statistics}. In the most common case of $p_{i}\ll1$, the result is
\begin{equation}
P_{i}(p_{i})=P_{i}(0)e^{-Np_{i}[1+\lambda(E_{i}-\Eav)]},
\label{eq:probability distribution}
\end{equation}
\begin{equation}
 \langle p_{i}\rangle =\frac{1}{N[1+\lambda(E_{i}-\Eav)]}.
\label{eq:paverage}
\end{equation}
The parameter $\lambda$ has the meaning of the inverse Hilbert space temperature. It can be calculated numerically by substituting the averaged probabilities (\ref{eq:paverage}) into the constraint (\ref{Eav}) on the energy expectation value:
\begin{equation}
 \sum_{i=1}^{N}(E_{i}-\Eav)\langle p_{i}\rangle=0.
\label{eq:lambda}
\end{equation}

In Refs. \cite{Fine-09-statistics,Fresch-10B} it was also shown that, for macroscopic systems consisting of $N_s$ weakly interacting constituents and for  $\Eav < 0 - O(E_1/\sqrt(N_s)$, the QMC ensemble implies condensation into the ground state, which means that the ground state acquires large average occupation 
\begin{equation}
 \langle p_{1} \rangle \approx \frac{\Eav}{E_{1}},
\label{eq:condensation}
\end{equation}
(typically of the order of 1), while all other eigenstates have occupations of the order of $1/N$ given by formula (\ref{eq:paverage}).
The probability distribution $P_1(p_1)$ in this case is not given by formula 
(\ref{eq:probability distribution}). Instead, it is narrowly peaked around $p_1 = \langle p_{1} \rangle$.

For other kinds of energy spectra or for non-macroscopic systems, the QMC condensation is also possible, but it may be accompanied by the anomalously large occupations of a few lowest energy levels instead of just one level\cite{Fine-10,Hantschel-11}.

The above condensation together with the very slowly decaying tail of form (\ref{eq:paverage})  leads to a non-thermal density matrix for a small subsystem within a macroscopic system\cite{Fine-09-statistics,Fresch-10A,Fresch-10B}.

In the following, the occupation numbers $\langle p_{i} \rangle$ generated as a result of numerically simulated quenches  are compared with prediction (\ref{eq:paverage}) for the QMC ensemble. In order to obtain $\lambda$, we combine Eq.~(\ref{eq:paverage}) and Eq.~(\ref{eq:lambda}). The result is
\begin{align}
 \sum_{i=1}^{N}\frac{E_{i}-\Eav}{N[1+\lambda (E_{i}-\Eav)]}=0.
\label{eq:lambda2}
\end{align}
The value of $\lambda$ is then obtained numerically by locating the zero of the LHS of Eq. \eqref{eq:lambda2} closest to $\lambda=0$.

Below, we also compare the quench-generated ensembles with the  canonical thermal ensemble corresponding to temperature $T$, which is chosen such that the thermal average energy is equal to $\Eav$:
\begin{align}
 \Eav= \frac{1}{Z} \sum_{i=1}^{N}E_{i}e^{-\frac{E_{i}}{T}}.
\end{align}
Here $Z$ is the partition function. In principle, there is no fundamental reason for the thermal ensemble to emerge, but such a comparison still provides a useful perspective. 

For the random-matrix quench considered in this work, there is an argument why the QMC ensemble may be expected to emerge. Namely, the transformation between the eigenbasis of the Hamiltonian $\Ho$ and the eigenbasis of the Hamiltonian $\Ham$ can be reasonably expected to be close to a random rotation in the Hilbert space. If it were indeed a random rotation, and then, in addition, the constraint on the energy expectation value were imposed, then the resulting set of states $\phi_k$ would indeed represent the QMC ensemble. 

In general, however, for physically realizable quenches, the Hamiltonian  $\Ham$ has the character of a band matrix in the basis of the Hamiltonian $\Ho$. Therefore, the transformation between the eigenbasis of $\Ham$ and the eigenbasis of $\Ho$ is not supposed to have the character of a random rotation. Yet, if the system is not too large, then one can expect that the QMC ensemble would provide at least a rough guide to the behavior of the function $\langle p \rangle (E)$. In particular, it was shown in Ref.~\cite{Ji-11} that multiple small quenches applied to finite clusters of spins 1/2 lead to QMC-like ensembles. If a random matrix quench also produces a QMC-like ensemble, it may imply that the effect of  multiple small physical quenches may be represented by a single random-matrix quench. 

\section{Random Matrix Hamiltonians}\label{sec:RandomMatrices}

All numerical results to be presented in this section are obtained for random matrices of size $N\times N=4096\times 4096$ with all diagonal elements and a fraction 30/4096 of the off-diagonal elements being different from zero.  The non-zero entries are to be selected from different ensembles.

\begin{figure} \setlength{\unitlength}{0.1cm}
%=======================================================================

\begin{picture}(100, 50)
{ 
\put(0, 0){ \epsfxsize= 2.9in \epsfbox{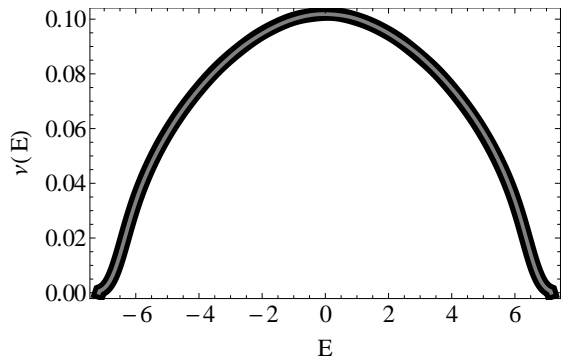} }
}
\end{picture} 
%============== 
\caption{Normalized densities of states $\nu(E)$ for two $4096 \times 4096$ sparse random matrix Hamiltonians $\Ham$ described in the text: thick black line --- all non-zero elements are real number randomly selected from the interval $[-1,1]$; thin grey line --- the same Hamiltonian but with one special diagonal element equal to -4.
} 
\label{fig:DOS} 
\end{figure}

\begin{figure} \setlength{\unitlength}{0.1cm}
%=======================================================================

\begin{picture}(100, 50)
{ 
\put(0, 0){ \epsfxsize= 2.9in \epsfbox{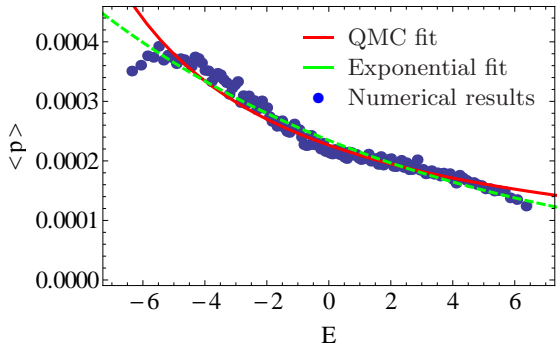} }
\put(47, 40){\small QMC fit}
\put(47, 36){\small Exponential fit}
\put(47, 32){\small Numerical results}
\color{red}
\thicklines
\put(40, 41){ \line(1,0){4} }
\color{qmcgreen}
\put(40, 37){ \line(1,0){4} }
\color{blue}
\put(42, 33){ \circle*{1.5} }
}
\end{picture} 
%============== 
\caption{(Color online) Occupations of eigenstates after a random matrix quench with all non-zero matrix elements in $\Ham$ being real and selected randomly from the interval $[-1,1]$. The energy expectation value for the initial state is $\Eav=-0.9$.   Numerical results represent 100 independent realizations of random matrices. The occupations of eigenstates $\langle p \rangle$ are averaged within energy bins, each containing $\Nb = 100 \cdot 32=3200$ states. The bin averages are computed according to formulas (\ref{eq:averaging1}) and (\ref{eq:averaging2}).  The QMC prediction (\ref{eq:paverage}) is represented by red
solid line. The exponential thermal distribution with same average energy is represented by the dashed green line.
} 
\label{fig:previous results}
\end{figure}

The random matrix quench for the matrix of the above size with all non-zero entries being real and chosen randomly from interval $[-1,1]$  was first investigated (without calling it a "quench") in Ref. \cite{Fine-09-statistics}. The density of states for this random matrix is shown in Fig.~\ref{fig:DOS}.   The resulting occupations of eigenstates appeared to be very well describable by the QMC formula (\ref{eq:paverage}). However, as we illustrate below, the very good agreement reported in Ref. \cite{Fine-09-statistics} was somewhat accidental. As we discovered later, the statistical fluctuations between different initial states and different random matrix realizations are larger than the visible statistical noise for the occupations of energy bins.  More complete averaging confirms overall satisfactory agreement, but also  reveals noticeable discrepancies between the numerically generated ensembles and the QMC prediction, in particular, near the edges of the energy spectrum. For example, Fig.~\ref{fig:previous results}, shows these discrepancies for one of the values of $\Eav$ used in Ref. \cite{Fine-09-statistics}.  We return to the implications of Fig.~\ref{fig:previous results} later in subsection~\ref{localization}. 

Here we focus on the fact that the interesting regime of possible QMC condensation could not be reached with the above random matrix ensemble because of the following constraint: The QMC condensation regime requires $\Eav$ to be sufficiently close to the the ground state energy. In this case, however, the lowest possible energy expectation value $\Eav$ for the initial state $\phi_k$ is the smallest possible value of the diagonal element of $\Ham$, i.e. $-1$, which is not close enough to the ground state energy equal to about $-6$ [see Fig. \ref{fig:DOS}]. Making the random matrix larger or less sparse would only lower the ground state energy without changing the minimum value of $\Eav$. On the other hand, making the matrix more sparse can indeed bring the ground state energy closer to the minimum value of $\Eav$, but, at the same time, the basis states $\phi_k$ become too correlated with the eigenstates $\psi_i$. In the limit, when all off-diagonal elements are equal to zero, the eigenstates of $\Ho$ trivially coincide with the eigenstates of $\Ham$. Therefore, on the approach to this limit, the random rotation argument given in Section \ref{sec:QMC} becomes increasingly inadequate, and the ensemble generated by the quench becomes very different from QMC.

The main part of the next subsection introduces and investigates the new ensemble of random matrices containing one special diagonal element selected differently than the rest. This allows us to realize the QMC condensation regime. The idea is that, on the one hand, only one special diagonal element does not strongly influence the shape of the energy spectrum, while, on the other hand, this element gives us a predetermined value of $\Eav$ which can be chosen much closer to the ground-state energy of $\Ham$.  We further investigate ensembles with more than one special diagonal element. Finally, we will show that the result of Ref.~\cite{Fine-09-statistics} extends to a complex-valued random matrix selected from the Gaussian unitary ensemble.

\subsection{Results}

\subsubsection{Ensembles with one or more special diagonal elements}

In this subsection, we consider sparse random matrices, where all non-zero off-diagonal elements and all but one diagonal elements are chosen randomly from a uniform distribution in the interval [-1,1]. A single diagonal element is assigned a value smaller than $-1$. We then choose the basis state $\phi_k$ corresponding to this special element as the initial state before the quench, i.e. $\Eav$ becomes smaller than -1.

\begin{figure} \setlength{\unitlength}{0.1cm}
%=======================================================================

\begin{picture}(100, 197)
{ 
\put(0, 193){(a)}
\put(0, 144){(b)}
\put(0, 147){ \epsfxsize= 2.9in \epsfbox{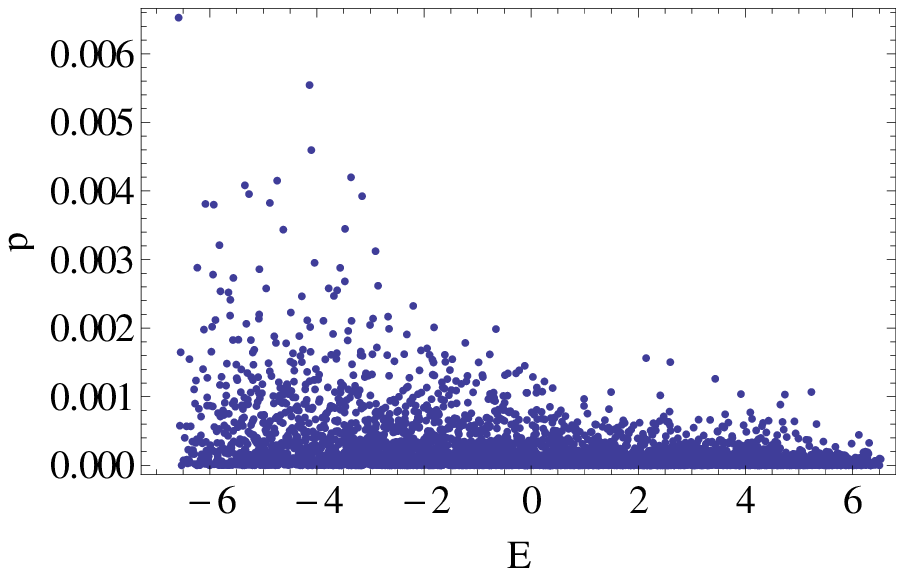} }
\put(0, 98){ \epsfxsize= 2.9in \epsfbox{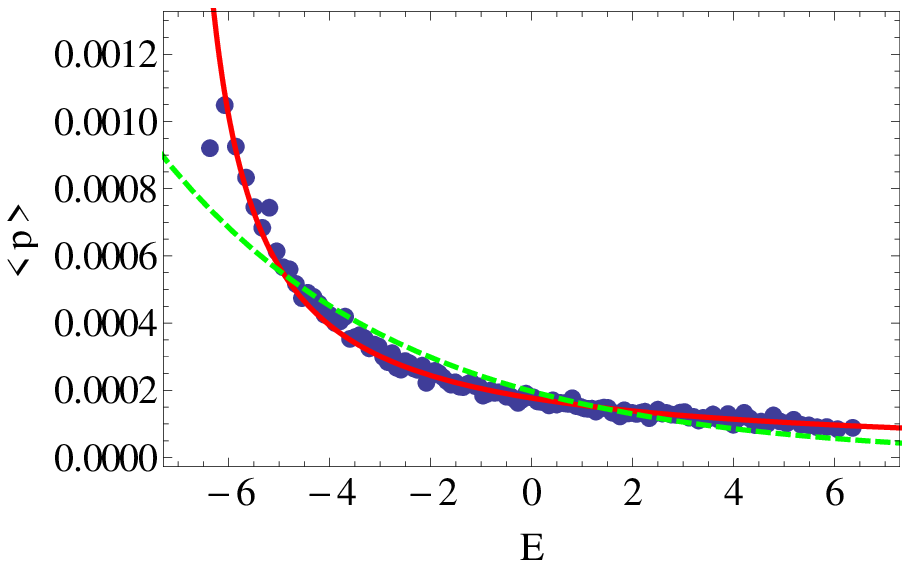} }
\put(47, 134){\small QMC fit}
\put(47, 130){\small Exponential fit}
\put(47, 126){\small Numerical results}
\color{red}
\thicklines
\put(40, 135){ \line(1,0){4} }
\color{qmcgreen}
\put(40, 131){ \line(1,0){4} }
\color{blue}
\put(42, 127){ \circle*{1.5} }
\color{black}
\put(0, 96){(c)}
\put(0, 47){(d)}
\put(0, 0){ \epsfxsize= 2.9in \epsfbox{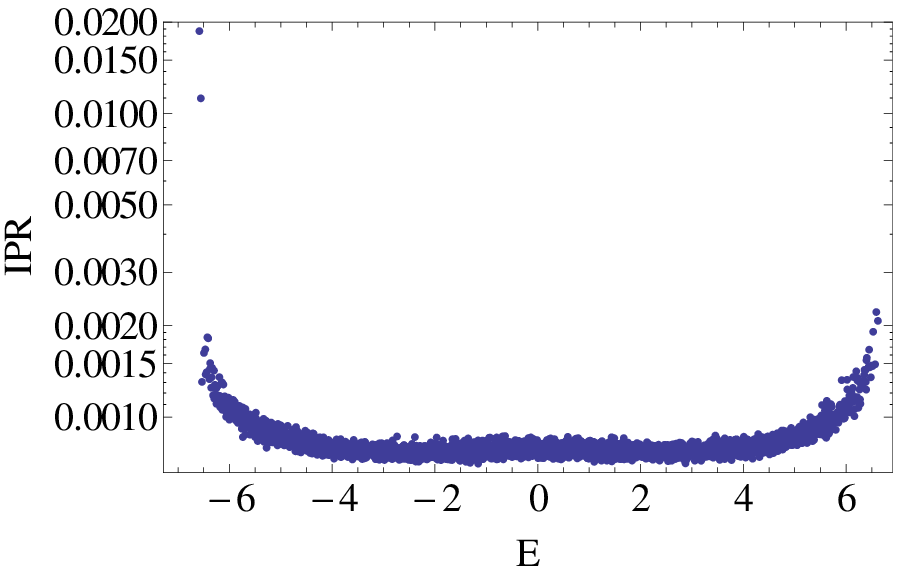} }
\put(0, 49){ \epsfxsize= 2.9in \epsfbox{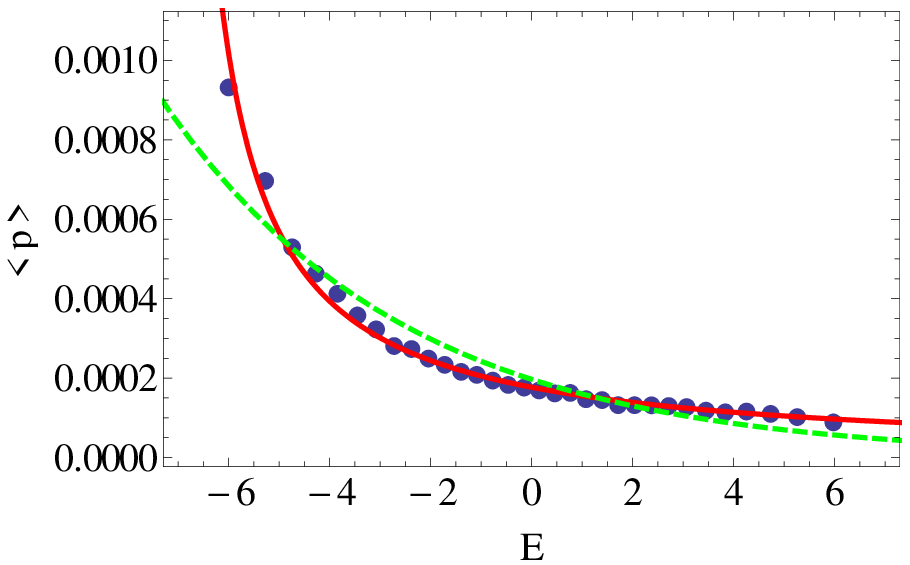} }
\put(47, 85){\small QMC fit}
\put(47, 81){\small Exponential fit}
\put(47, 77){\small Numerical results}
\color{red}
\thicklines
\put(40, 86){ \line(1,0){4} }
\color{qmcgreen}
\put(40, 82){ \line(1,0){4} }
\color{blue}
\put(42, 78){ \circle*{1.5} }
}
\end{picture} 
%============== 
\caption{(Color online) Occupations of eigenstates after a random matrix quench for the sparse Hamiltonian $\Ham$ with one special diagonal element as described in the text. The energy expectation value for the initial state is $\Eav=-2$. (a) Unaveraged occupations of eigenstates $p$ for a single realization of the quench. (b-c) Bin-averaged occupations of eigenstates for 20 realizations of the quench with bin size equal to $\Nb = 20 \cdot 32=640$ in (b) and $\Nb = 20 \cdot 128=2560$ in (c). The bin averages are computed according to formulas (\ref{eq:averaging1}) and (\ref{eq:averaging2}).  The QMC prediction (\ref{eq:paverage}) is represented by red
solid line. The exponential thermal distribution with same average energy is represented by the dashed green line. (d) Inverse participation ratios of eigenstates given by Eq.(\ref{eq:IPR}) for a single realization of $\Ham$.
} 
\label{fig:main result} 
\end{figure}

We start with the example of $\Eav = -2$ illustrated in detail in Fig.~\ref{fig:main result}. Figure~\ref{fig:main result}(a) shows unaveraged occupations of eigenstates after the quench for a single realization of the random matrix Hamiltonian $\Ham$. Figures~\ref{fig:main result}(b) and (c) show the occupations of eigenstates averaged over 20 realizations of random matrices. All pairs $\{E_i, p_i \}$ from all 20 realizations are combined in an energy-ordered list, and then divided into the energy bins, with each bin containing $\Nb$ entries.  
Fixing the same $\Nb$ for each bin implies that the bins acquire varying energy widths and may become particularly broad near the edges of the energy spectrum. Here and everywhere below,  the average characteristics for the $j$-th energy bin are computed as follows:
\begin{align}
 \bar E_{j}&=\frac{\sum \limits_{j\text{-th bin}}p_{i}E_{i}}{\sum \limits_{j\text{-th bin}}p_{i}}, \label{eq:averaging1}\\
\langle p \rangle_{j}&=\frac{1}{N_b}\sum_{j\text{-th bin}}p_{i}. \label{eq:averaging2}
\end{align}
The difference between Figs.~\ref{fig:main result}(b) and (c) is that, in the former case, the bin size is smaller, $\Nb = 20\cdot32=640 $, while in the later case it is larger, $\Nb = 20\cdot128=2560$.
Finally Fig.~\ref{fig:main result}(d) shows the inverse participation ratio (IPR) for the expansion of eigenstates $\psi_i$ in terms of the basis states $\phi_k$ --- see Eq.(\ref{eq:IPR}).

As seen in Fig.~\ref{fig:main result}(b) the bin-averaged occupations for all but the lowest energy bin follow the QMC statistics very well. It is also very clear that the occupations of the eigenstates do not decrease exponentially as one would expect from a thermal distribution but rather decrease according to the power law of the QMC ensemble. In Fig.~\ref{fig:main result}(c), the discrepancy for the lowest energy bin was eliminated by choosing the bin size such that the lowest energy bin includes all the eigenstates with anomalously high inverse participation ratio.  This aspect is to be discussed further in subsection~\ref{localization}.

\begin{figure} \setlength{\unitlength}{0.1cm}
%=======================================================================

\begin{picture}(100, 101)
{ 
\put(0, 97){(a)}
\put(0, 47){(b)}
\put(0, 50){ \epsfxsize= 2.9in \epsfbox{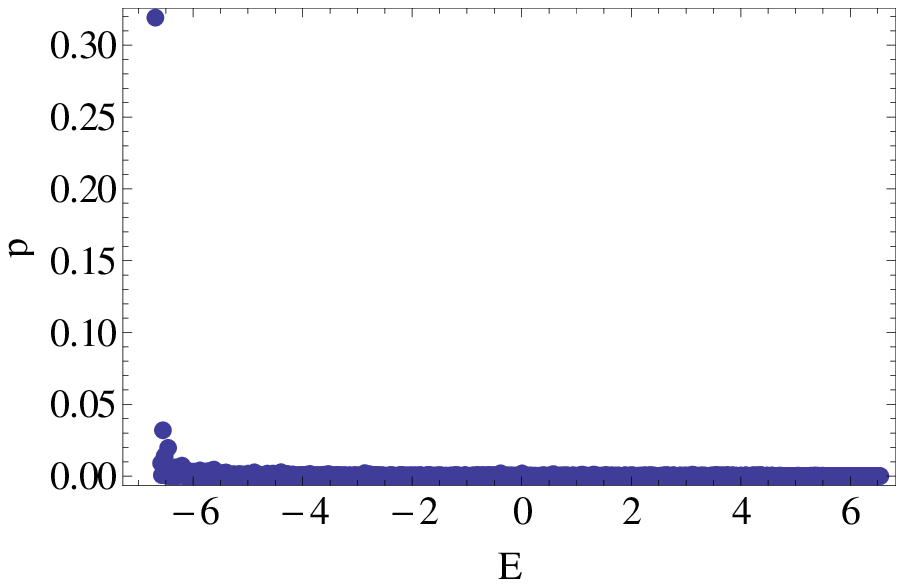} }
\put(0, 0){ \epsfxsize= 2.9in \epsfbox{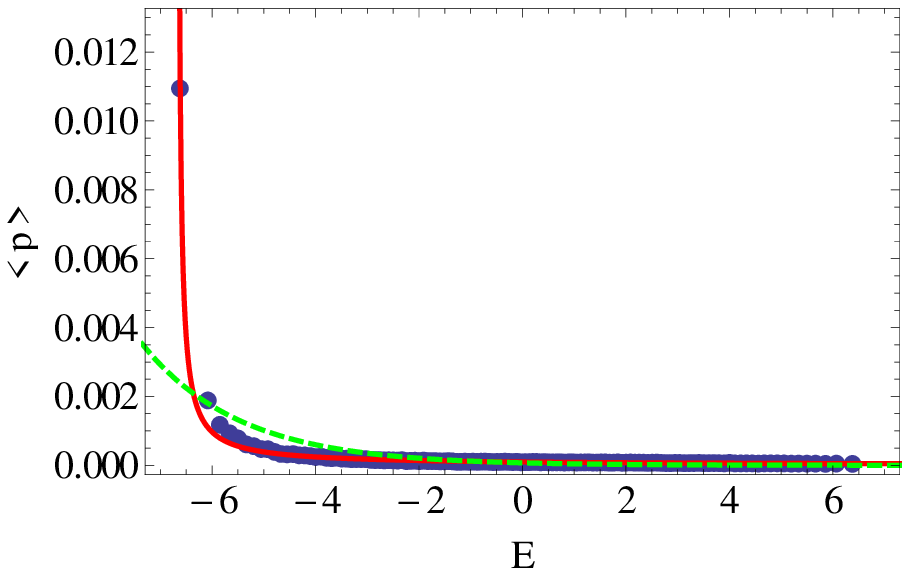} }
\put(24, 15){ \epsfxsize= 1.9in \epsfbox{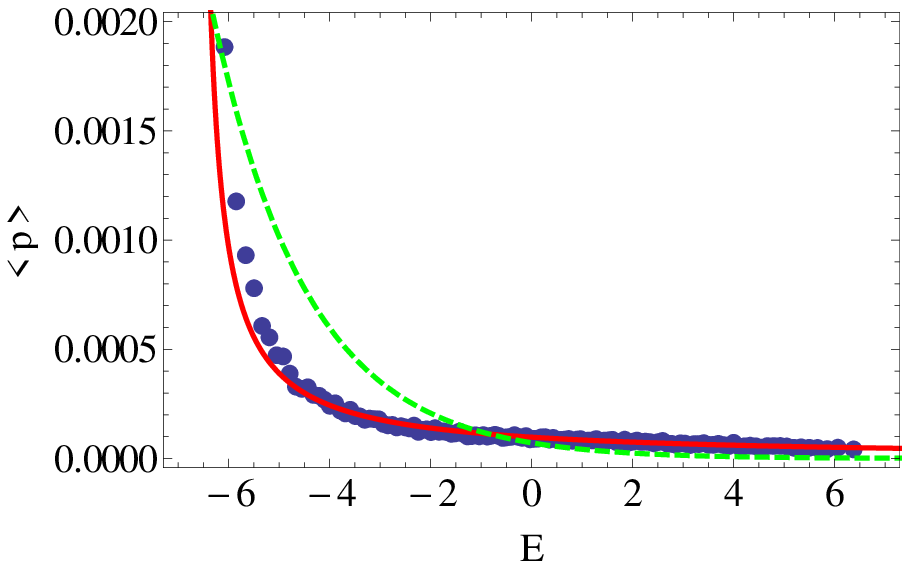} }
}
\end{picture} 
%============== 
\caption{(Color online) Frames (a) and (b) have the same kind of content and the same plot legends as Figs.~\ref{fig:main result}(a) and (b), respectively, but with the special diagonal element giving $\Eav=-4$. The large occupation of the ground state exhibited in frame (a) indicates the QMC condensation regime. The inset magnifies the small-$\langle p \rangle$ part of the main plot in (b). The density of states for this case is shown in Fig.~\ref{fig:DOS}.
} 
\label{fig:condensation} 
\end{figure}

Figure~\ref{fig:main result}(a) indicates that a quantum quench with $\Eav = -2$ does not exhibit the condensation into a single eigenstate, yet. However, by reducing the value of the special diagonal element further, the QMC condensation can indeed be reached. In Fig.~\ref{fig:condensation} we present the result of the quench for the same kind of random matrix but with the special diagonal element equal to -4. Fig.~\ref{fig:condensation}(a) shows that the ground state has occupation comparable to 1, while the occupations of all other states are very small and, as shown in Fig.~\ref{fig:condensation}(b), accurately describable by the QMC statistics. As shown in Fig.~\ref{fig:DOS}, the density of states in this case is still nearly the same as in the case of random matrix without a special diagonal element.

\begin{figure} \setlength{\unitlength}{0.1cm}
%=======================================================================

\begin{picture}(100, 50)
{ 
\put(0, 0){ \epsfxsize= 2.9in \epsfbox{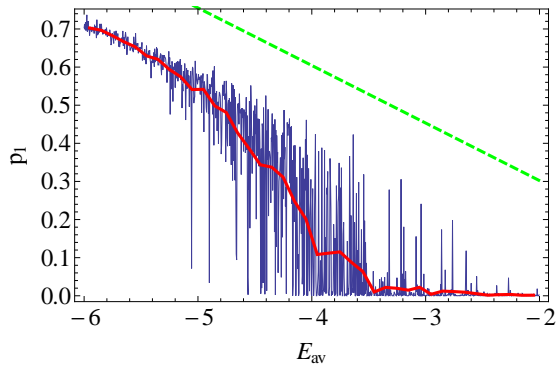} }
}
\end{picture} 
%============== 
\caption{(Color online) The onset of condensation into the ground state as a function of $\Eav$ for the random matrix quench with one special diagonal element. Noisy blue line - occupation of the ground state $p_1$ for a single realization of random matrix with a given value of the special diagonal element.  The solid red line shows $p_1$ averaged over the random realizations. Dashed green line represents the prediction (\ref{eq:condensation}) for the QMC ensemble in a macroscopic system.
} 
\label{fig:condensation2}
\end{figure}

In Fig. \ref{fig:condensation2}, we plot the occupation $p_{1}$ of the ground state as a function of $\Eav$. The numerically generated values of $p_{1}$ strongly fluctuate as a function of $\Eav$, because each plot point was generated with different realization of random matrix. Overall, however, the averaged occupation  $\langle p_{1} \rangle(\Eav)$ shows quite a sharp transition to the condensation regime at an average energy of about -3.5. In all cases, the occupation of the lowest eigenstate lies significantly below the value given by Eq. (\ref{eq:condensation}) for a QMC ensemble in a macroscopic system with short-range interactions.

\begin{figure} \setlength{\unitlength}{0.1cm}
%=======================================================================
\begin{picture}(100, 100)
{ 
\put(0, 97){(a)}
\put(0, 47){(b)}
\put(0, 50){ \epsfxsize= 2.9in \epsfbox{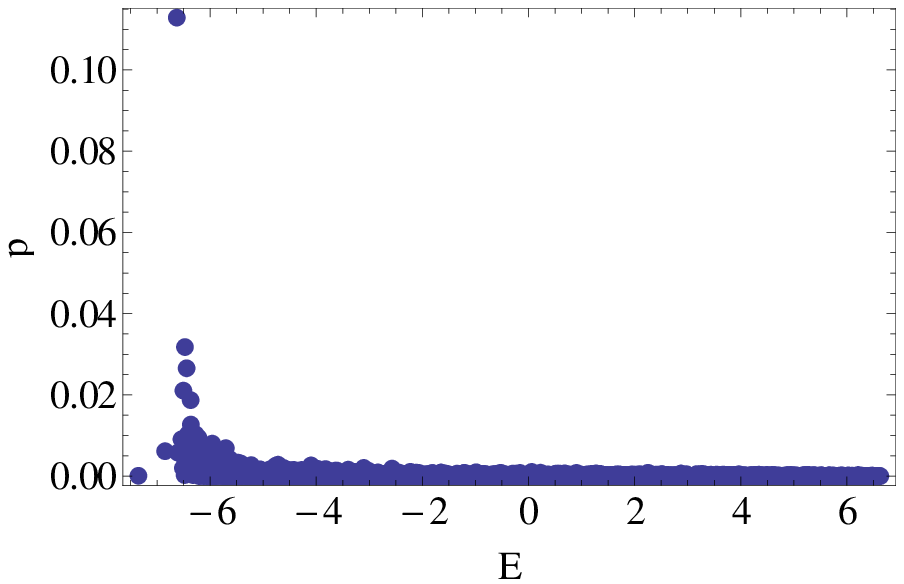} }
\put(0, 0){ \epsfxsize= 2.9in \epsfbox{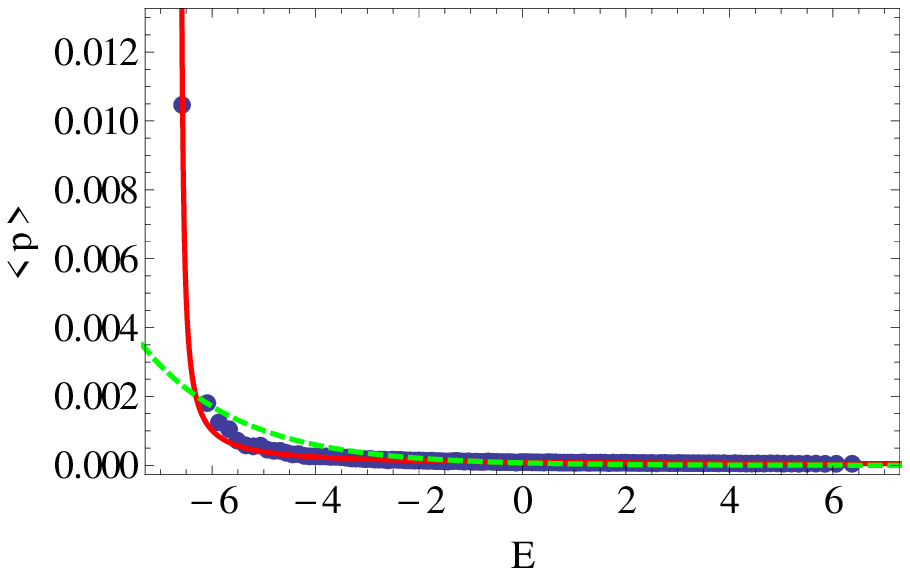} }
\put(24, 15){ \epsfxsize= 1.9in \epsfbox{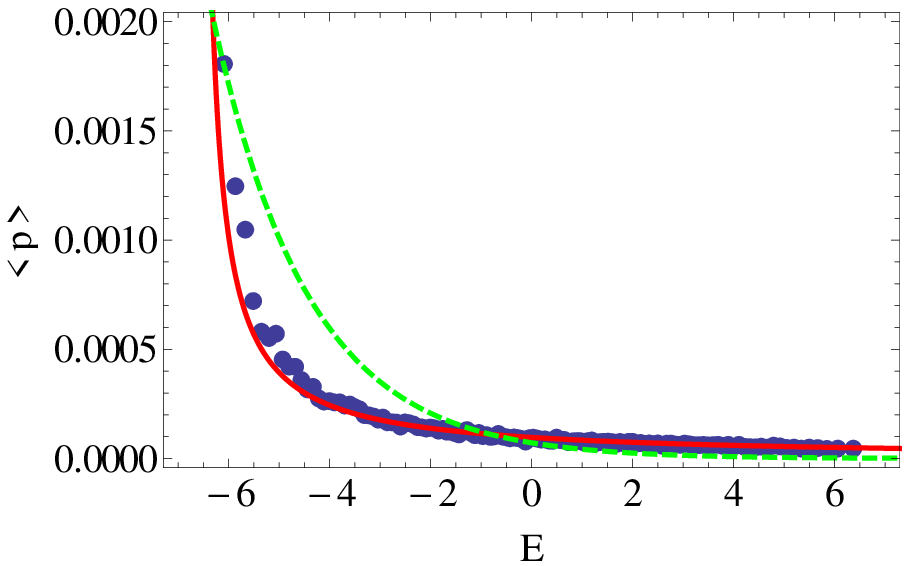} }
}
\end{picture} 
%============== 
\caption{Same as Fig.~\ref{fig:condensation} but for the random matrix quench with three special diagonal elements -5, -4.5, -4 and $\Eav=-4$.
} 
\label{fig:special_elements} 
\end{figure}

In order to investigate the condensation regime further we introduce not one but three special diagonal elements with values -4, -4.5 and -5, and then choose the basis state corresponding to the diagonal element equal to -4 as the initial state before the quench. The results are presented in Fig.~\ref{fig:special_elements}. We observe that two eigenstates with the lowest and the second lowest energy are almost unoccupied. The third eigenstate from below has large occupation comparable to one, i. e. we observe condensation but not into the lowest energy state. The averaged occupations of the remaining eigenstates agree very accurately with the QMC formula. In the present case, choosing the basis states corresponding to the diagonal elements -5 or -4.5 as the initial states would lead to the condensation into the lowest or the second lowest eigenstates, respectively. We note here that condensation into the second or the third lowest energy state contradicts to the QMC statistics always predicting the condensation into the lowest energy state. We discuss this disagreement in subsection~\ref{localization}. 

\begin{figure} \setlength{\unitlength}{0.1cm}
%=======================================================================

\begin{picture}(100, 48)
{ 
\put(0, 0){ \epsfxsize= 2.9in \epsfbox{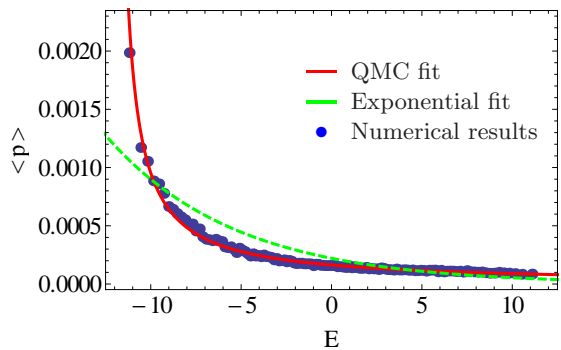} }
\put(47, 36){\small QMC fit}
\put(47, 32){\small Exponential fit}
\put(47, 28){\small Numerical results}
\color{red}
\thicklines
\put(40, 37){ \line(1,0){4} }
\color{qmcgreen}
\put(40, 33){ \line(1,0){4} }
\color{blue}
\put(42, 29){ \circle*{1.5} }
}
\end{picture} 
%============== 
\caption{Same as Fig.~\ref{fig:condensation}(b) but for the random matrix quench with non-zero entries in $\Ham$ selected according to the prescription for the Gaussian unitary ensemble as described in the text and with $\Eav = -4$.
} 
\label{fig:gaussian} 
\end{figure}

\subsubsection{Sparse Gaussian unitary ensemble}

Here we investigate the case, when $\Ham$ is a random matrix of the same size and with the degree of sparseness as before but with the non-zero matrix elements selected according to the prescription for the Gaussian unitary ensemble: the real and the imaginary part of each non-zero off-diagonal element is chosen randomly from the Gaussian distribution with variance $\sigma=1/\sqrt{2}$; the diagonal elements are real and chosen from the Gaussian distribution with variance $\sigma=1$. A diagonal element equal to -4 is added, and the corresponding basis state is expanded in terms of eigenstates. The result presented in Fig. \ref{fig:gaussian} exhibits a very good agreement with formula (\ref{eq:paverage}) for the QMC statistics. In the present case, the condensation regime was not reached for most random matrix realizations, although a small minority of the realizations did exhibit it, implying the condensation precursor behavior similar to the one observed in Fig.~\ref{fig:condensation2} above $\Eav = -3.5$.

\subsection{Discussion: correlations, condensation and localization}
\label{localization}

The numerical result presented in Fig.~\ref{fig:previous results} indicates that for modest values of $\Eav$ random matrix quenches lead to the ensembles that are approximately describable by the QMC statistics but with noticeable deviations from the prediction by Eq. \eqref{eq:paverage}. The QMC formulas \eqref{eq:probability distribution} and \eqref{eq:paverage} correspond to the limit of $N \gg 1 $, which should indeed be appropriate for $4096$ states involved. As discussed below, the deviations near the edges of the energy spectrum, can be attributed to the localization properties of the eigenstates involved. More difficult, however, is to understand the discrepancy in the middle range between $E= -4$ and $E=4$. Here we observe, that the deviation between the numerical result and formula (\ref{eq:paverage})  looks somewhat similar to the deviation between the same formula and the direct simulations of the QMC ensemble for the quantum systems with the number of levels of the order of 10 \cite{Fine-10, Hantschel-11}. The results of those QMC simulations are not describable by formula \eqref{eq:paverage} because of the finite-$N$ effects. We therefore suspect, that since the random matrix is sparse with each line having only about 30 non-zero off-diagonal elements, the basis states may form resonantly coupled groups. The number of these groups may indeed be of the order of 10.  Within each group, the basis states delocalize into a band-like structure, which is correlated in the sense that it is not obtained by a random rotation of the basis states belonging to the group. In the next order, these groups randomly couple to each other thereby forming the resulting eigenstates. If true, such a scenario suggests, that the
eigenstates of the random-matrix Hamiltonians considered emerge not through a truly random rotation of the entire 4096-dimensional Hilbert space, but rather  through a rotation of a smaller subspace thereby leading to the deviations from the QMC predictions. The fact that band-like correlations lead to deviations from the QMC statistics will be further substantiated in Section \ref{sec:DisorderedPotentials} on disordered potentials.  

Figure~\ref{fig:condensation2} illustrates that random matrices with one special diagonal element exhibit a rather sharp transition to QMC-like condensation as a function of the special element.
In general, we observe in Figs. \ref{fig:main result}, \ref{fig:condensation},
\ref{fig:special_elements} and \ref{fig:gaussian}, that the agreement with the QMC prediction improves as this prediction approaches the condensation regime. Once the condensation regime is reached, as in Figs. \ref{fig:condensation} and
\ref{fig:special_elements}, the occupations of non-condensed states are very well describable by the QMC formula. 

The QMC ensemble implies condensation into the lowest energy state for macroscopic systems with finite Hilbert space, reasonably short range of interactions and almost any average energy per particle (in the spectral range of the system). The density of states for the total energy of such a system has a narrow Gaussian maximum in the middle  with half-width, which is much smaller than  the entire range of the energy spectrum. In contrast, the density of states of random matrices considered in this work are approximately semicircle, i.e. the maximum in the middle is very broad. In this case, the QMC condensation is still possible, but only for the values of $\Eav$ sufficiently far from the middle of the spectrum. Another consequence of the absence of the narrow peak is the nonapplicability of formula (\ref{eq:condensation}) for the condensed fraction (see Fig. \ref{fig:condensation2}). 

As shown in Fig.~\ref{fig:special_elements}, when more than one special diagonal element is introduced, the condensation occurs not necessarily into the ground state of the system, but possibly into a higher energy eigenstate. This circumstance together with the fact that in Fig. \ref{fig:main result}(b) the deviations from the QMC statistics occur for the bin of the lowest-energy states indicates that the phenomenon of localization may be involved here. 
The QMC ensemble requires a fair sampling of the Hilbert space. This implies a broad participation of eigenstates $\psi_i$ in a given basis state $\phi_k$. But if some of the eigenstates are localized among very few basis states, they will not be present in most of the quench-induced ensembles considered in this paper. This localization property can be characterized by the inverse participation ratio (IPR) of basis states $\phi_{k}$ in a given energy eigenstate $\psi_{i}$:
\begin{equation}
 \text{IPR}_{i}=\sum_{k=1}^{N}|\langle \psi_{i}|\phi_{k}\rangle|^{4}
\label{eq:IPR}
\end{equation}
IPR$_i$ of the order of $1/N$ indicates that eigenstate $\psi_i$ is delocalized;
IPR$_i \gg 1/N$ indicates that the eigenstate is localized.
We note here that the phenomenon of QMC condensation, if observed in the random matrix quench, indicates that the lowest-energy eigenstate is strongly localized. 

In general, however, it is not just one eigenstate, but a group of eigenstates near both edges of the spectrum that are localized. 
A plot of the inverse participation ratio of eigenstates for a matrix ensemble with special element equal to -2 is presented in Fig.~\ref{fig:main result}(d).
The plot shows that  states at very low energies are localized in the sense of having a rather high inverse participation ratio. Therefore, it is not surprising that the QMC statistics fails to describe their individual occupations.  

We define the ``mobility edge'' through the critical value IPR$_{cr}$ of the inverse participation ratio as
\begin{align}
\text{IPR}_{\hbox{\scriptsize cr}}= 1.5 \cdot \text{IPR}_{\hbox{\scriptsize typ}},
\end{align}
where IPR$_{\hbox{\scriptsize typ}}$ is the average value of the inverse participation ratio of an eigenstate in the middle of the spectrum. We call an eigenstate localized, if its inverse participation ratio is larger than IPR$_{\hbox{\scriptsize cr}}$. 

The idea is now to put all eigenstates below the mobility edge into a single energy bin.
If the QMC statistics works above the mobility edge, and all states below the mobility edge belong to a single bin, then the QMC prediction will be valid for the lowest energy bin as well. It is guaranteed by the normalization condition 
\begin{equation}
 \langle p_{1} \rangle = 1 - \sum_{i=2}^{N}\langle p_{i} \rangle.
\label{eq:p1_normalization}
\end{equation}
The effectiveness of the above binning procedure can be observed by comparing Fig.~\ref{fig:main result}(b) and (c). In Fig.~\ref{fig:main result}(b) the lowest energy bin is not large enough. Hence it does not include all localized eigenstates and exhibits clear deviation from the QMC prediction. In Fig.~\ref{fig:main result}(c),  the bin size was increased, so that all eigenstates below the mobility edge belong to the first energy bin, which then exhibits a very good agreement with the QMC prediction.

To summarize, our findings suggest that the QMC statistics works best, when the combined occupation of all eigenstates below the mobility edge is significant and all these states are grouped into a single energy bin. 

In the above discussion we neglected the existence of the second mobility edge near the top of the energy eigenspectrum.  In general, the occupations of eigenstates above the high-energy mobility edge also tend to deviate from the QMC statistics, but these occupations are small when the condensation regime on the other end of the energy spectrum is approached.

\section{Disordered Lattices}
\label{sec:DisorderedPotentials}

In this section, we consider one particle in a single-band lattice. The Hamiltonian before the quench has disordered on-site energies $\varepsilon_i$ and no hopping elements:
\begin{equation}
\Ho = \sum_{i=1}^{N}\varepsilon_{i}a_{i}^{\dagger}a_{i}.
\end{equation}
Here $a_{i}$ is the particle annihilation operator for the $i$th site.
The particle before the quench occupies the $k$th site and has energy $\Eav = \varepsilon_{k}$.   
The quench consists of switching-on the nearest-neighbor hopping term
\begin{equation}
 \Hp = \sum_{\langle i,j \rangle} \left( t_{ij} a_{i}^{\dagger}a_{j} + h.c. \right)
\end{equation}
where $t_{ij}$ are the hopping elements, which are also disordered. The notation $\langle i,j \rangle$ implies the summation over nearest neighbors.

\begin{figure} \setlength{\unitlength}{0.1cm}
%=======================================================================

\begin{picture}(100, 167)
{ 
\put(0, 163){(a)}
\put(0, 108){(b)}
\put(0, 53){(c)}
\put(33, 157){ {\textbf{3D fcc lattice}} }
\put(0, 110){ \epsfxsize= 2.9in \epsfbox{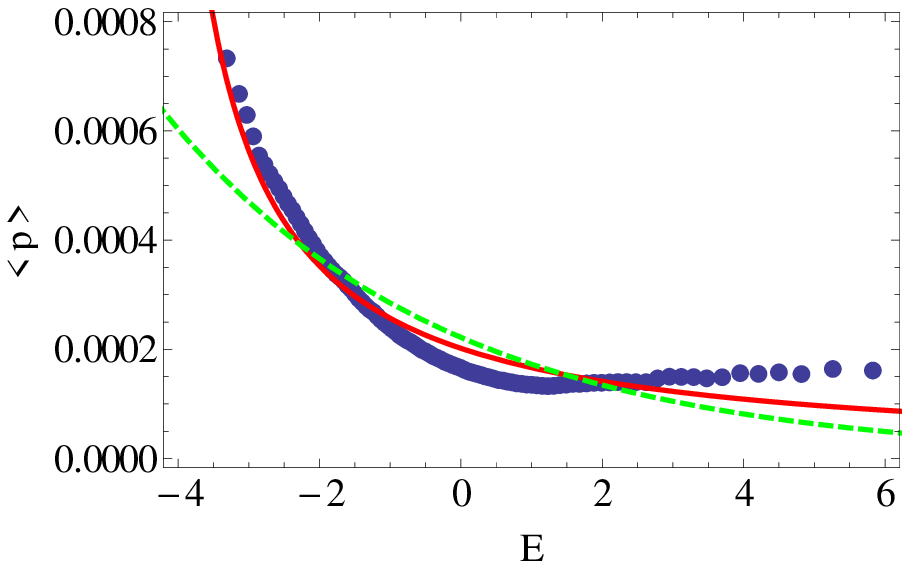} }
\put(25, 102){ {\textbf{3D simple cubic lattice}} }
\put(0, 55){ \epsfxsize= 2.9in \epsfbox{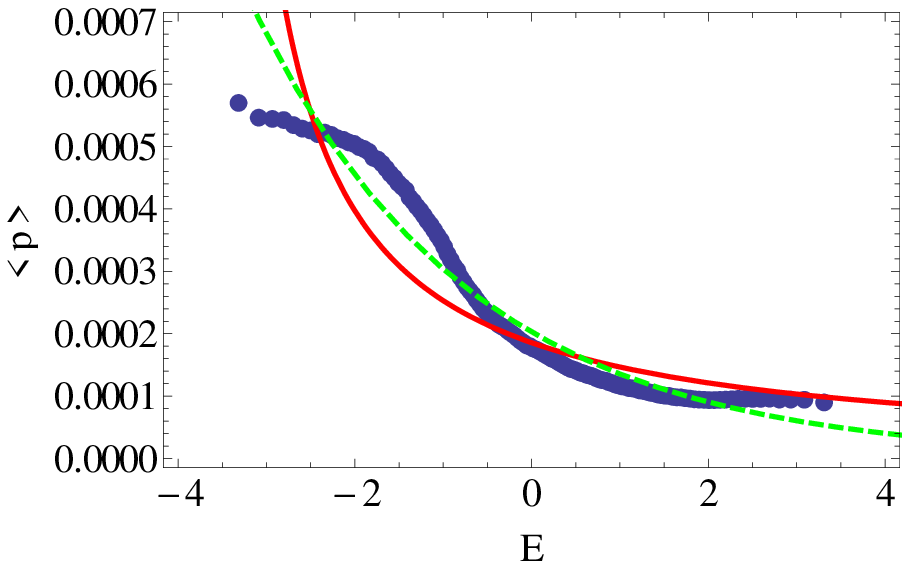} }
\put(35, 47){ {\textbf{1D lattice}} }
\put(0, 0){ \epsfxsize= 2.9in \epsfbox{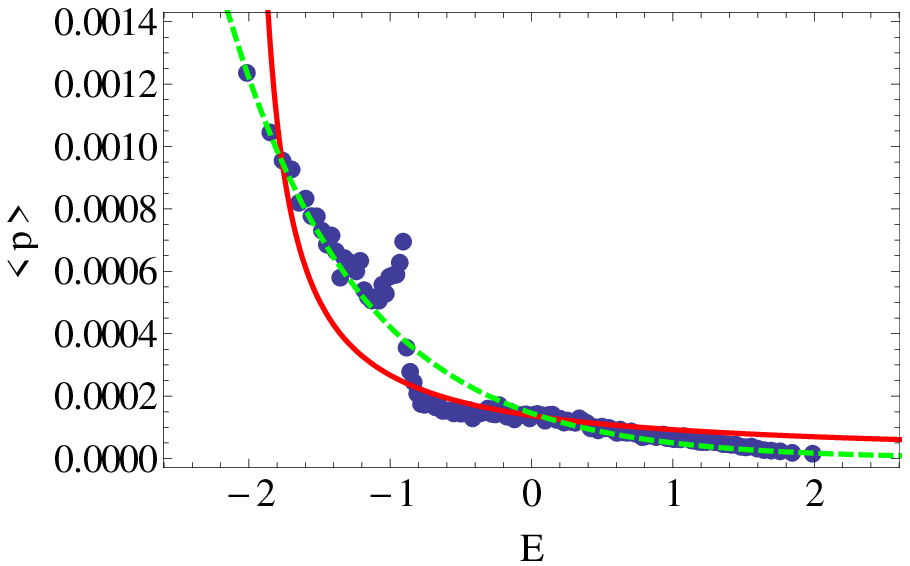} }
\put(34, 20){ \epsfxsize= 1.5in \epsfbox{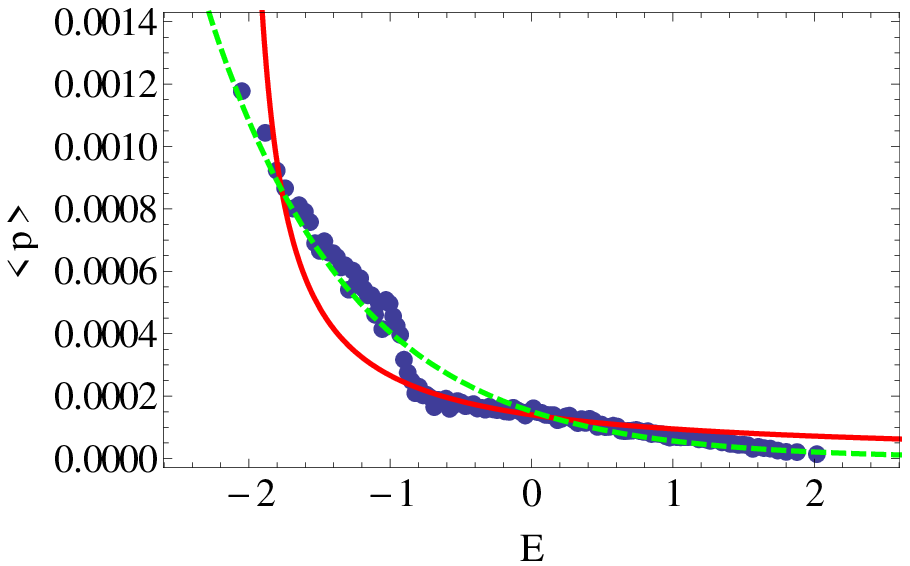} }
\put(47, 146){\small QMC fit}
\put(47, 142){\small Exponential fit}
\put(47, 138){\small Numerical results}
\color{red}
\thicklines
\put(40, 147){ \line(1,0){4} }
\color{qmcgreen}
\put(40, 143){ \line(1,0){4} }
\color{blue}
\put(42, 139){ \circle*{1.5} }
}
\end{picture} 
%============== 
\caption{(Color online) Quantum quenches for different disordered lattices with random on-site energies $\epsilon_i$ selected from the interval $[-1,1]$ and real random hopping elements $t_{ij}$ selected from the interval $[0,1]$. The type of lattice is indicated above each plot. The inset in (c) shows the results for the 1D lattice with hopping elements selected from the interval $[0.1, 1]$. In all cases, the energy expectation value of the initial states is $\Eav = -0.9$.  The numerical results represent 600 realizations of each disordered lattice. To improve the statistics further, 20  basis states with energy expectation value closest to -0.9 were included.  
The plots show the bin-averaged occupations of eigenstates $\langle p \rangle$ with bin sizes $\Nb$ equal to $600\cdot20\cdot40=480000$ in (a) and $600\cdot20\cdot32=384000$ in (b) and (c). The solid red line represents the QMC prediction by Eq. (\ref{eq:paverage}). The dashed green line represents the exponential thermal distribution with same average energy.  
} 
\label{fig:disordered potentials} 
\end{figure}

The full Hamiltonian ${\cal H} = \Ho + \Hp $ is sparse, but, unlike the case of random matrices, it has non-zero off-diagonal elements located at positions that are not random. 

Physically, this situation may be realized with disordered potentials, when initially the potential barriers between the wells are so large that the hopping can be neglected, and then the barriers are suddenly lowered, so that the particle can hop. Or, alternatively, a single particle is injected into the $k$th site of the lattice describable by the full Hamiltonian ${\cal H} = \Ho + \Hp$.
Both settings should be realizable in the experiments on optical lattices. They are also conceptually relevant to the experiments with disordered speckle potentials \cite{Schwartz-07,Aspect-08,DeMarco-11,Jendrzejewski-12} and quasi-periodic potentials \cite{Roati-08} that reported the evidence of Anderson localization \cite{Anderson-58}.  

\subsection{Results and discussion}
\subsubsection{Disordered lattices with real-valued hopping elements}

In this part, we investigate three lattices: 

(i) three dimensional (3D) face centered cubic (fcc) lattice with 4000 sites; 

(ii) 3D simple cubic lattice with 4096 sites; and 

(iii) one-dimensional (1D) chain with 4096 lattice sites.

\noindent
For each lattice, we choose the on-site energies $\epsilon_i$ randomly from a uniform distribution in the interval [-1,1] and hopping elements $t_{ij}$ randomly from a uniform distribution in the interval [0,1]. The latter choice is motivated by the fact that for real potentials with deep enough wells, all hopping elements should be positive. The initial site is characterized by the on-site energy $\Eav \approx -0.9$. The results are presented in Fig. \ref{fig:disordered potentials}.

For all three lattices, both the QMC formula \eqref{eq:paverage} and the exponential distribution can describe the occupations of eigenstates only very roughly, with all numerically generated distributions exhibiting large deviations. These deviations may be related to band-like correlations mentioned in subsection~\ref{localization}, because the densities of states for all three lattices considered have traces of the band structures associated with the limit of no onsite disorder and translationally invariant values of $t_{ij}$. In the translationally-invariant case, the occupation of each eigenstate (Bloch state) would have equal value, which is clearly different from the QMC prediction.   On the other hand, we note, that the deviations from the QMC statistics do not appear to have systematic character, which implies that the approximation based on the QMC ensemble is, probably, the best one can do without performing a system-specific analysis. In particular, the QMC statistics can be reasonably used to estimate the probability of a particle being found above or below the mobility edge after the particle is injected into a site with a given value of the on-site energy $\Eav$.

{\it A priori}, we expected that the result for the fcc lattice would be closest to the random matrix case because of the largest number of nearest neighbors (equal to 12) and thus closest to the QMC prediction. As seen in Fig. \ref{fig:disordered potentials}(a), this expectation was indeed confirmed. 

On the opposite end is the result shown in Fig. \ref{fig:disordered potentials}(c) for 1D lattice, which exhibits sharp edge and a peak, both close to $\Eav$. The above peak can be understood by observing that each site of the 1D lattice has only two nearest neighbors. The corresponding hopping elements are chosen randomly from the interval [0,1], which means that there is a non-zero chance for both of them to be very close to zero. In such a case, the initial state is close to a strongly localized eigenstate, which means that the occupation of a single eigenstate is comparable to 1. The inset of Fig. \ref{fig:disordered potentials}(c) shows the result for a simulation with hopping elements chosen from the interval [0.1,1], i.e. the possibility of the near-zero hopping elements is excluded. The peak at the average energy vanishes, as expected from the above qualitative picture. For the other two lattices in Fig. \ref{fig:disordered potentials}, the probability of all hopping elements around a given site to be close to zero is suppressed due to higher numbers of nearest neighbors. Hence the peak disappears.

\subsubsection{Disordered lattices with complex-valued hopping elements}

\begin{figure}[t] \setlength{\unitlength}{0.1cm}
%=======================================================================

\begin{picture}(100, 50)
{ 
\put(0, 0){ \epsfxsize= 2.9in \epsfbox{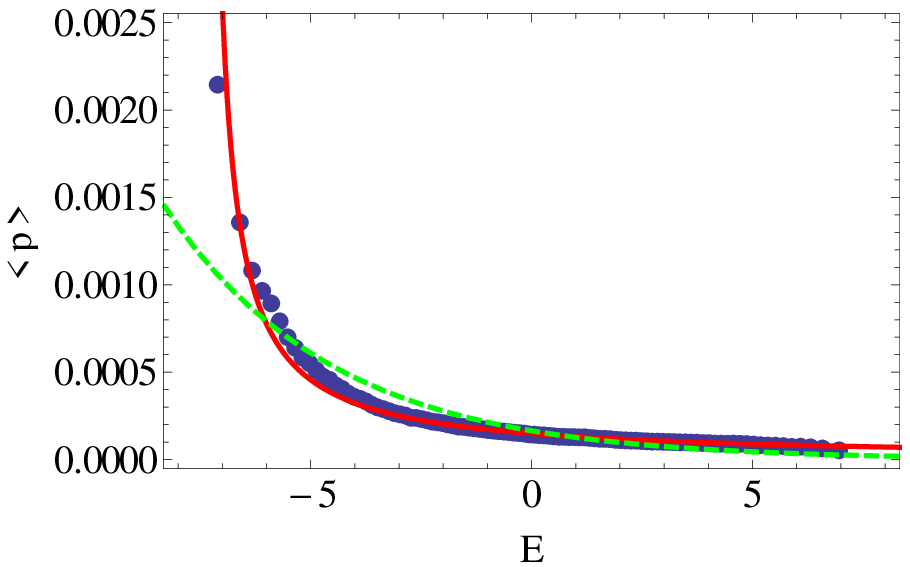} }
\put(24, 15){ \epsfxsize= 1.9in \epsfbox{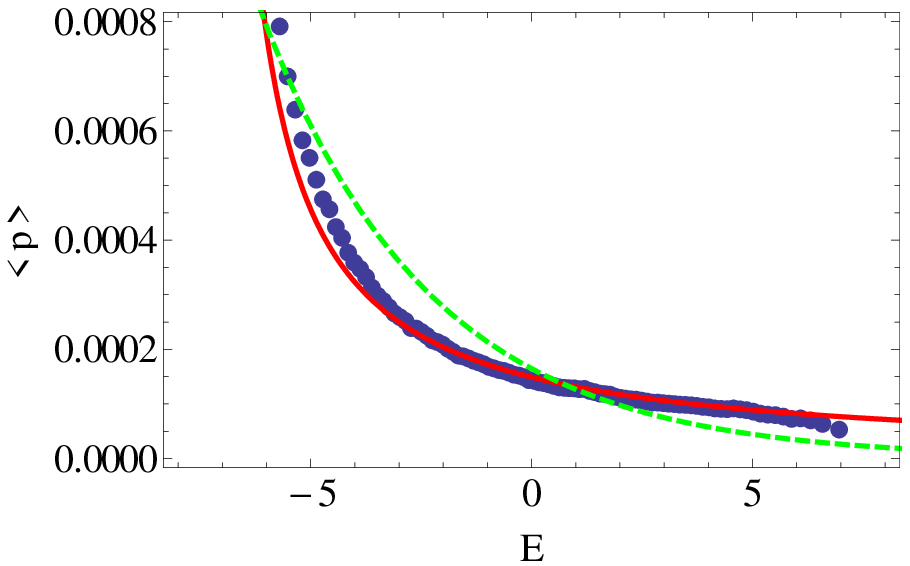} }
}
\end{picture} 
%============== 
\caption{(Color online) Same as Fig.~\ref{fig:disordered potentials}(a) but with the on-site energies and the hopping elements selected according to the prescription for the Gaussian unitary ensemble as described in the text. Energy expectation value of the initial state is $\Eav=-3$.  600 independent realizations of the disordered fcc lattice were used. The size of the energy bin is $\Nb = 600\cdot40=24000$. The inset magnifies the small-$\langle p \rangle$ part of the main frame.
} 
\label{fig:disordered potentials complex} 
\end{figure}

In experiments with ultra cold atoms on optical lattices, it may be possible to break the time-reversibility by simulating the effect of a staggered magnetic field  acting on a charged particle \cite{Hemmerich-10}. In this case, the hopping elements $t_{ij}$ are complex. We investigated this case by studying the fcc lattice with complex-valued hopping elements.

In this case, we chose the on-site energies from a Gaussian distribution with variance $\sigma = 1$, and both Re$(t_{ij})$ and Im$(t_{ij})$ (independently) from a Gaussian distribution with variance $\sigma = 1/\sqrt{2}$. Initially, the particle is located on a site with $\Eav = -3$.  The result is presented in Fig.~\ref{fig:disordered potentials complex}. In contrast to the results for the lattices with real and positive hopping elements, the agreement of the bin-averaged occupations of eigenstates with the QMC formula \eqref{eq:paverage} is very good. This is consistent with the expectation that, in the present case, band-like correlations are destroyed more efficiently due to the random phases of the hopping elements. Also $\Eav$ in this case is closer to the QMC condensation threshold.

\section{Conclusions}

We have investigated quantum ensembles emerging as a result of off-diagonal quenches of sparse random-matrix Hamiltonians, and
disordered potentials.
The resulting ensembles are controlled by the energy expectation value
of the initial quantum state. We have introduced a new class of random-matrix Hamiltonians with one special diagonal element and shown that quantum quenches with these Hamiltonians exhibit transition to
the  QMC condensation. In general,
random-matrix eigenstates close the edges of the energy
spectra tend to be more localized in the original basis, and hence the
occupations of these eigenstates follow the QMC statistics not so well.
However, if a mobility edge for the eigenstates is introduced, and all
states below the mobility edge are grouped  into a single energy bin,
then the QMC statistics is recovered. In the case of disordered
potentials, our simulations indicate that the resulting ensembles, in
general, do not exhibit quantitative universality. Nevertheless, the QMC
statistics can still serve at least as a crude approximation. The quality of this approximation improves significantly when the number of the nearest neighbors on the lattice is increased, or when the time-reversibility is broken.

%\bibliography{matri}

\end{document}